# Sub-wavelength localization of near-fields in coupled metallic spheres for single emitter polarization analysis

Po-Nan Li[1], Hsiu-Hao Tsao[1], Jer-Shing Huang[2], and Chen-Bin Huang[1,*]

[1]*Institute of Photonics Technologies, National Tsing Hua University, Hsinchu 30013, Taiwan*
[2]*Department of Chemistry, National Tsing Hua University, Hsinchu 30013, Taiwan*
*\*Corresponding author: robin@ee.nthu.edu.tw*

We numerically demonstrate selective near-field localization determined by the polarization state of a single emitter coupled to plasmonic nano-cluster. Seven gold nanospheres are carefully arranged such that up to ten polarization states of the single emitter, including linear, circular, and elliptical polarizations, can be distinguished via the distinct field localization in four gaps. The ability to transform polarization state into field spatial localization may find application for single emitter polarization analysis. © 2011 Optical Society of America

The generation and control of surface plasmons have enabled the realization of optical devices with subwavelength footprints, such as optical antennas [1,2] and spasers [3,4]. Through geometrical design, the characteristics of surface plasmon resonance can be tailored and the corresponding local fields react sensitively to the phase and polarization of the incoming electromagnetic waves. This opens the path to spatial and temporal control of the optical near-fields that may find important applications in future plasmonic optical nanocircuitry.

To localize optical fields, one may exploit the near-field interference by manipulating the coherent properties of the far-field excitation [5]. Recently, several remarkable reports have demonstrated the control of propagating direction and the localization of optical near-fields [6-8]. On the other hand, the ability to manipulate the polarization of single emitters or that of local fields is also important. To this aim, linear optical antennas were used to control the directivity and polarization of single quantum emitters [9].

However, linear nanoantennas only enhance one field component and are therefore limited to linear polarization control. Cross optical nanoantennas were proposed as plasmonic polarization controllers to overcome this problem and extend the ability of polarization control [10,11]. Exploiting the geometry-dependent resonance of LSP, cross nanoantennas with an asymmetric gap were used to spatially localize optical near-fields according to the frequencies such that sub-wavelength spatial resolution was achieved and the antennas worked as nano-colorsorters [12,13]. Plasmonic spirals, in which focused or defocused near-fields are generated according to the handedness of the far-field illumination, have been demonstrated to function as polarization analyzers [14,15]. These works demonstrated either the ability to control the polarization or the spatial localization of near-fields via far-field illumination. However, the capability to achieve spatial control of the near-field distribution by a local source and thus the capacity to analyze the polarization states, including the handedness of the local sources, remain unexplored. The ability to access the polarization information of single quantum emitters and local sources is important since they play a pivotal role in the envisaged optically integrated circuits and quantum communications [16].

In this Letter, we numerically demonstrate multiple selective localization of optical near-fields excited by a single-frequency emitter with various polarization states. Through proper geometrical design of seven coupled gold nanospheres, up to ten polarization states of the local source are successfully converted into distinguishable LSP spatial distributions. We therefore obtain the polarization state, including linear, circular, and elliptical polarizations, of the local emitter. The unique polarization to LSP spatial localization mapping within the four air gaps formed by the seven nanospheres can find immediate application as true nanoscopic polarization analyzer for near-field optical sources. Methods to implement our design are discussed.

Figure 1(a) shows the schematic of our plasmonic structure. Seven identical gold nanospheres with radius $r$=80 nm, center-to-center pitch $d$=190 nm are positioned on the x-y plane. The spheres are surrounded by vacuum and are correspondingly numbered. Spheres 2 and 3 both make an angle of $\theta$=45° with respect to sphere 1. A pointwise electrical field source is placed 20 nm above the surface of sphere 1 ($z$=100 nm) to excite the plasmonic structure. The simulations are performed with the three-dimensional finite-difference time-domain (FDTD) method [17]. The simulation volume is large enough to avoid nonphysical absorption of the near fields by the boundaries and an uniform mesh step $((5\ nm)^3)$ is used such that satisfactory calculation accuracy is obtained with reasonable speed and memory consumption. The optical source wavelength is 1550 nm. The dielectric function of gold is modeled using Drude-Lorentz function to match the experiments [18]. The instantaneous field intensity within the four air gaps labeled A, B, C, and D are recorded for quantitative analyses.

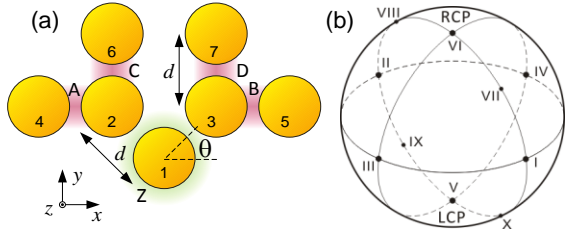

Fig. 1. (Color online) (a) Seven coupled gold nano-spheres. Intensities within air gap A, B, C and D are used for polarization analyses. (b) The ten polarization states considered are shown on a Poincaré sphere. RCP and LCP: right- and left-hand circular polarization.

The ten input polarizations considered within this Letter are labeled on a Poincaré sphere as depicted in Fig. 1(b). States I, II, III and IV denote 45°, 135°, 0°, and 90° linearly polarized inputs, respectively. States V and VI represent left- and right-hand circularly polarization inputs. States VII, VIII, IX, and X stand for the four elliptical polarizations on the meridian, with a relative phase delay between the x- and y-polarized electrical fields of 45°, 135°, 225° and 315°, respectively.

Since every two coupled spheres work as an optical antenna [2], they show polarization-dependent response to the excitation source. Under the current geometry, only longitudinal mode (fields polarized along the interparticle line) can couple between spheres [19]. This fact makes LSP excitation in our structure sensitive to the optical source polarization. For example, when a 45° linearly polarized source is used, surface plasmons excited on sphere 1 is longitudinal with respect to the sphere 1-3 chain, but transverse with respect to sphere 1-2 chain. Thus energy coupling from spheres 1 to 3 is much more efficient than that between spheres 1 and 2. The surface plasmons energy is then evenly coupled onto spheres 5 and 7 since sphere 3 serves as a T-junction for a particle-chain plasmonic waveguide [6,7]. Enhanced intensities are thus observed within gaps B and D, which we define as LSP selective state I. Following the same analogy, LSP state II, where intensities are localized within gaps A and C, is selectively excited using a 135° linearly polarized source. Such enhanced field localization (or "Hot spot") can be understood as a result of near-field interference of the plasmon resonance on the coupled spheres similarly found in random composites [5]. Alternatively, it can also be seen as the standing-wave pattern of the propagating guided modes in a branched waveguide with finite length, similar to a Fabry-Pérot resonator [2].

These anticipations are verified through the FDTD simulations. Figure 2(a) shows the steady-state near-field intensity distributions of the structure when excited by a 45° linearly polarized source. LSP are spatially localized within gaps B (between spheres 3 and 5) and D (between spheres 3 and 7). Figure 2(b) shows the result excited under a 135° linearly polarized input source, and LSP are selectively localized within gaps A and C. Note the quantitative agreement that the intensities of the localized fields in two sets of gaps (B and D for 45°; A and C for 135°) have equal value. Figures 2(c) and 2(d) show the near-field intensity map under 0° (state III) and 90° (state IV) linearly polarized source excitation, respectively.

Our design shows clear and distinct selective field localization, but is much more simplified and works even with single-frequency source, as compared to that demonstrated in Refs. [6-8], where complicated waveforms and/or genetic algorithm were required to reach a 2-state selectivity. In fact, our simple design is capable of achieving multiple selective field localizations corresponding to different source polarization states, including circular and elliptical polarizations.

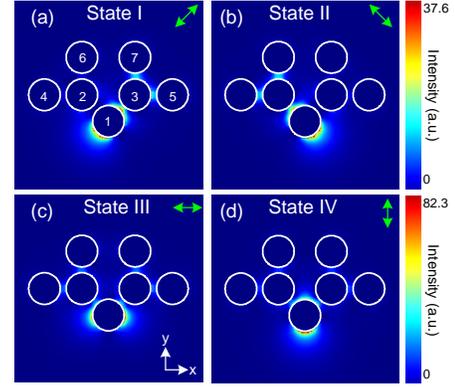

Fig. 2. (Color online) LSP intensity distributions excited by (a) 45°, (b) 135°, (c) 0°, and (d) 90° linearly polarized source.

For a nanoscopic polarization analyzer, the main challenge is in discerning the handedness of the inputs. Figures 3(a) and 3(b) show the FDTD results when left- (state V) and right-hand (state VI) circularly polarized inputs are used to excite our structure, respectively. Under circularly polarized excitations, the LSP spatial distributions are asymmetric to the y-axis and the handedness of the input can be directly identified: LSP are localized within gaps A and D in Fig. 3(a), but confined within gaps B and C in Fig. 3(b). Circular polarizations can be decomposed into two orthogonal linearly polarized inputs with equal amplitudes but a ±90° phase shift. We examined the FDTD results and the fields within the two LSP-filled gaps are temporally 90° degrees out of phase. This is attributed to the coherent interactions of the LSP fields within the gaps under these two phase-delayed linearly polarized inputs [5].

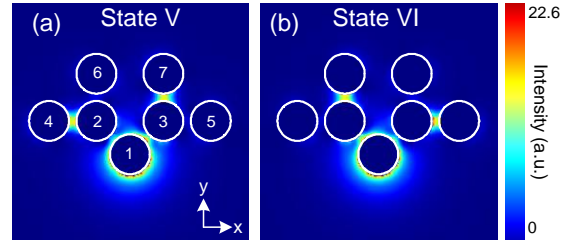

Fig. 3. (Color online) LSP intensity distribution under (a) left- and (b) right-handed circularly polarized inputs.

Results shown in Fig. 3 provide convincing evidence that selectively excited LSP states in our designed plasmonic structure can effectively function as a true nanoscopic polarization analyzer. Compared to that reported in Refs. [14,15], where scanning and fitting of the near-field intensity patterns to Bessel functions are required, our structure is capable of distinguishing the two circularly polarized inputs directly through the selective LSP-filled gap locations. Such ability to analyze complex polarization state and the handedness of local sources is of great interest in circular dichroism determination of single molecules [20,21] and provide possible solutions for high resolution spatial mapping of chiral polymer films.

Lastly, we extend the demonstration to elliptically polarized inputs. The resulting LSP spatial distributions are anticipated as the intermixture of those observed under the adjacent linear and circular polarized inputs. For example, Fig. 4(a) shows the LSP distribution excited by input state VII as defined in Fig. 1(b), revealing the resemblance of both selective LSP states I (Fig. 2(a)) and VI (Fig. 3(b)). This new selective state results from the superposition of the near-fields of LSP states I and VI, giving three LSP-filled gaps (B, C, and D) and one LSP-void gap (A). Figures 4(c-d) show the selective LSP intensity distributions corresponding to input excitation state of VIII, IX and X, respectively. The results are similar to that observed in Fig. 4(a): one LSP-void gap (correspondingly labeled in the figures) accompanied by three LSP-filled gaps. Through the determination of the LSP-void gap location, elliptical input polarizations can thus be analyzed.

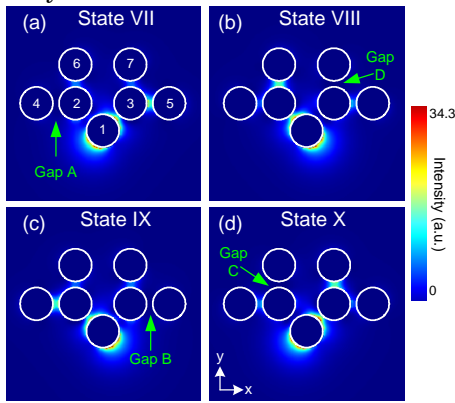

Fig. 4. (Color online) LSP intensity distributions under the four elliptically polarized inputs as shown in Fig. 1(b). The input polarization states and the LSP-void gap locations are correspondingly labeled.

In summary, four air gaps formed by seven gold nanospheres are designed to selectively localize enhanced near-fields and serves as polarization analyzer for a local emitter. Such simple design translates the polarization state of a single local emitter onto distinguishable selective localization of enhanced near-fields. Up to 10 polarization states, including circular and elliptical polarizations can be distinguished. The selective field localization is achieved without manipulation of complex broadband excitation. The ability to distinguish other points on the Poincaré sphere should be possible if quantitative measurement on the field intensity are taken into account and is under further investigations.

Our proposed scheme may be realized by incorporating quantum emitters onto the plasmonic structure, and the resulting near-fields measured through near-field scanning optical microscopy, photon localization microscopy [13], or two-photon photoemission electron microscopy [8]. Under the current geometry, a spatial resolution of 20 nm or better in the SNOM measurement should be adequate for the polarization read-out. As one of the future research goals, it would be interesting if the selective LSP excitation can be confined within one specific air gap. This could be made possible by optical arbitrary wave generations [22], where another degree of freedom in the temporal domain can be exploited. Other designs, including metallic nano-discs or cubes in different lattice geometry, may also find practical applications provided the design rule is well understood.

This work was supported by National Science Council in Taiwan under grants NSC 97-2112-M-007-025-MY3, NSC 99-2120-M-007-010 and NSC 99-2113-M-007-020-MY2.

# References with title